\pdfoutput=1
\documentclass[a4paper]{jpconf}
\usepackage{amsfonts,palatino,amsthm,latexsym,mathrsfs}
\usepackage[fleqn]{amsmath}
\usepackage{amssymb}
\usepackage{dsfont}
\usepackage{epsfig}
\usepackage{textcomp}
\usepackage{longtable} 
\usepackage{hyperref} 
\usepackage{multirow}
\usepackage{enumerate} 
\usepackage{color} 
\usepackage[margin=2.5cm]{geometry} 

\DeclareMathAlphabet{\mathpzc}{OT1}{pzc}{m}{it}
\def\eali\end{align*}
\def\be{\begin{equation}}
\def\ee{\end{equation}}
\def\bea{\begin{eqnarray}}
\def\eea{\end{eqnarray}}
\def\bal{\begin{align}}
\newcommand{\eal}{\end{align}}
\def\ble{\begin{flalign}}
\newcommand{\ele}{\end{flalign}}
\def\ba{\begin{array}}
\def\ea{\end{array}}
\def\bali{\begin{align*}}
\def\nn{\nonumber \\}

\def\bite{\begin{itemize}}

\def\eat{\end{itemize}}
\def\begfig{\begin{figure}[h!]}
\def\endfig{\end{figure}}

\newcommand{\bwe}{\begin{widetext} \begin{eqnarray}}
\newcommand{\ewe}{\end{eqnarray}\end{widetext}}

\def\tfrac#1#2{{\textstyle{#1\over #2}}}
\def\half{\tfrac{1}{2}}

\def\quarter{\tfrac{1}{4}}

\def\del{\partial}
\def\deli{\del_i}
\def\delj{\del_j}

\def\delx0{\frac{\del}{\del x^0}}
\def\delx1{\frac{\del}{\del x^1}}
\def\delx2{\frac{\del}{\del x^2}}
\def\delx3{\frac{\del}{\del x^3}}

\def\pdif#1#2{\frac{\del #1}{\del #2}}

\def\dag{\dagger}

\def\dagg{{}^{\dag}}

\def\ket#1{\mid #1 {\cal{i}}}

\def\norm#1{\| #1 \|}

\def\vect{\overrightarrow}

\def\N{\ensuremath{\mathds{N}}}

\def\R{\ensuremath{\mathds{R}}}

\def\Lag{\ensuremath{\mathcal{L}}}

\def\half{\ensuremath{\frac{1}{2}}}
\def\so4{\ensuremath{\mathfrak{so(4)}}}
\def\su2{\ensuremath{\mathfrak{su(2)}}}
\def\SU2{\ensuremath{{SU(2)}}}

\def\gij{g_{ij}}

\def\rf#1{Eq.(\ref{#1})}

\def\ni{\noindent}

\def\bd{\bf \begin{definition}: \it}
\def\ed{\end{definition} \rm}
\def\blem{\bf \begin{lemma}: \it}
\def\elem{\end{lemma} \rm}
\def\bthe{\bf \begin{theorem}: \it}
\def\ethe{\end{theorem} \rm}
\def\bcor{\bf \begin{corollary}: \it}
\def\ecor{\end{corollary} \rm}
\def\bpro{\bf \begin{proof}: \rm}
\def\epro{\end{proof} \rm}

\def\undback16{& \!\!\!\!\!\!\!\!\!\!\!\!\!\!\!\!}

\def\+{\ensuremath{\ket{+}}}
\def\-{\ensuremath{\ket{-}}}

\def\Hs{\ensuremath{\mathcal{H}}}

\begin{document}
\title{Experimenting with Quantum Fields in Curved Spacetime in the Lab}

\author{Isabeau Pr\'emont-Schwarz}
\ead{isabeau@aei.mpg.de}
\address{Max-Planck-Institut f\"ur Gravitationsphysik, Albert Einstein Institute,\\
Am M\"uhlenberg 1, Golm, D-14476 Golm, Germany}
\begin{abstract}
In this paper we will investigate how one can create emergent curved spacetimes by locally tuning the coupling constants of condensed matter systems. In the continuum limit we thus obtain continuous effective quantum fields living on curved spacetimes. In particular, using Stingnet condensates we can obtain effective electromagnetism. We will show for example how we obtain quantum electrodynamics in a blackhole (Schwarzschild) spacetime.

\end{abstract}

\section{Introduction} \label{Intro}
Even though Hawking radiation\cite{hawkrad} is only a kinematical effect of QG, it is one of the very few predictions of an essentially quantum-gravitational nature which is generally believed to be true. In order to overcome our lack of access to highly curved chunks of spacetime, analogue models have been proposed where a curved spacetime is simulated by the effective equations of motion for light in dielectrics\cite{Felice}, of classical sound or surface waves in a liquid\cite{Unruh81}, of perturbations of a Bose-Einstein condensate\cite{Garay2, BLV1}, of perturbations of a super fluid\cite{JacVolvHe} or of electromagnetism in quantum dielectric\cite{LeonhardtSL}. 
In fact some experiments with analogue model black holes have already been made that claim to observe stimulated emission of Hawking radiation, though at a classical level \cite{Silkelab}.

What we would like to propose in this paper is a slightly different class of models which is obtained from a simple idea. Condensed matter systems can give rise to many types of quasiparticles, from Majorana fermions\cite{majoranareturns} to gauge bosons and anyons\cite{gaugebosqhall}, practically any type of matter one might want can be cooked up by some condensed matter system.  We will show in this paper that by varying the coupling constants, in space and in time\footnote{Space and time is, for the purpose of this sentence, understood to be the flat spacetime of the lab in which the condensed matter system is located.}, of a condensed matter system, we can simulate a curved spacetime for the quantum fields of the system. 

In the next section we investigate how a variable speed of light can be reinterpreted as a curved geometry with a constant speed of light. In section \ref{Emergence} we show how an effective metric emerges from a tweaked quantum system. In section \ref{QEDBH} we suggest a concrete realization of U(1)-Yang-Mills theory in a Schwarzschild black hole background. 

\section{Variable Speed of Light or Curved Geometry: two descriptions of the same thing.}\label{Varspeed}
Imagine a D-dimensional space $\R^D$ which we endow with the flat metric $\delta_{\mu \nu}$ on which a field can propagate information at a finite speed $c(\overrightarrow{x},\overrightarrow{u})$ which is anisotropic and inhomogeneous in the sense that it depends on the position in space, $\overrightarrow{x}$, and the direction $\overrightarrow{u}$, $\overrightarrow{u}$ being a unit D-vector (with respect to the flat spacial metric). If this field is all that exists in the space, it may be difficult to justify the use of the flat metric $\delta_{\mu\nu}$ as it is not based on anything observable. A more operational (and arguably physical) definition of distance should rely on observables in the theory. A particularly natural choice in our case is to measure distances by sending signals with the field\footnote{In the case of multiple fields, what follows will only work if all the fields propagate information at the same speed.}, thus relating elapsed time with distance. By definition then, we may define the speed of signalling to be $1$ at the expense of changing the metric. 
We thus started out with an flat spacetime metric $\eta$ but measuring distances by sending signals, we are compelled to change that metric to 
\begin{align}
 g(x,v,v): = \varphi(x)^2 (-(v^0)^2 + {k^2(x,\overrightarrow{v}/\norm{\overrightarrow{v}})}\overrightarrow{v}^2 ), \label{newmetric}
\end{align}
where $x:=(x^0,\overrightarrow{x})$, is the spacetime position, $v:=(v^0,\overrightarrow{v})$ is a tangent vector at $x$ and $k(x,\overrightarrow{v}/\norm{\overrightarrow{v}}) := \frac{1}{{c(x,\overrightarrow{v}/\norm{\overrightarrow{v}})}}$ is the inverse of the speed at location $x$ and in direction $\overrightarrow{v}$. Notice how we have an arbitrary local conformal factor $\varphi(x)\in\R_+$ which we cannot fix classically since to impose that the speed of signalling is one, we have the option of rescaling space by $x\rightarrow 1/v x$ or rescaling time by $t\rightarrow v t$ or a mixture of both; however, as we will see, quantum mechanics give us a scale ($\hbar$) against which to measure and fix this local conformal degree of freedom\footnote{For \rf{newmetric} to define a metric, certain conditions on the speed $c$ need to be respected, but as we will see, generic Lagrangian derived systems do indeed give rise to varying speeds which can be reinterpreted metrically. }

\section{Emergence of Spacetime}\label{Emergence}
The recipe to for cooking up matter fields on curved spacetimes in the lab is conceptually rather simple. The first step consists of choosing the type of matter wanted. Secondly, one finds a condensed matter system which whose collective degrees of freedom reproduce the type of matter sought after; fermionic fields, scalar fields, and the like all have condensed matter systems from which they can emerge. As explicitly shown in \cite{LRBint}, the speed of signals can be varied by varying the coupling constants. Thus the third step is to upgrade the coupling constants to functions of space and time, thereby making speed a function of spacetime. These functions can then be chosen so as to give the desired spacetime. Here is how it works in more detail. 

Suppose a quantum mechanical condensed matter system in D spacial dimensions.   In the limit where the lattice is very fine compared to measurements,i.e. the continuum limit, the discrete degrees of freedom can be approximated by continuous fields. Generically, the effective Lagrangian for the system, expressed in Lab spacetime coordinates, will be of the form
\begin{align}
\Lag(x) = -\left(\sum_a \half G_a^{ij} \deli \phi_a(x) \delj \phi_a(x) + \half M_a^2\phi_a^2(x)\right) -V(\phi_a(x))\label{genlag}
\end{align}
if in the continuum limit the degrees of freedom correspond to scalar fields (for example in the case of coupled quantum oscillators on a lattice).  In what follows we will assume that the continuum (or emergent) degrees of freedom are scalar fields, but the reasoning is the same for other types of fields. In fact, for the concrete example we give in section \ref{QEDBH} we have a spin-1 $U(1)$ gauge-field (``light'') describing the continuum limit of the degrees of freedom. $V$ is an function of the fields $\phi_a$ which is bounded from below (otherwise the resulting theory would be unstable and ill-defined). 

Without loss of generality we may suppose $G_a$ to be symmetric ($G_a^{ij}=G_a^{ji}$) since its antisymmetric part vanishes in any case in \rf{genlag} (for commutative momentum-spaces). The canonically conjugate momentum to the field $\phi_a(x)$ is then $\pi^a(x) = -G_a^{0 j}\delj \phi_a(x) $
Thus the Hamiltonian is 
\begin{align}
 \Hs =  \sum_a \pi^a\del_0\phi_a - \Lag =  \sum_a \left[\begin{array}{cc} \del_0\phi_a & \vect{\del}\phi_a \end{array}\right] \left[\begin{array}{cc} -G_a^{0 0} & \vect{0} \\ \vect{0} & G^s_a \end{array}\right] \left[\begin{array}{c} \del_0\phi_a \\ \vect{\del}\phi_a \end{array}\right] + \sum_a \half M_a^2\phi_a^2(x) + V(\phi_a(x)) , \label{Habibi}
\end{align}
where $G^s_a$ is the spacial part of $G_a$ (i.e. $G_a$ with the zeroth line and column removed). For the theory to be well defined, the Hamiltonian must be bounded from below, this implies that the matrices in \rf{Habibi} must be positive definite which in turns implies that $G_a^{0 0}<0$ and $G^s_a$ must be positive definite which means that the matrix $G_a$ must have Minkowskian, $(-,+, \ldots, +)$, signature.

If we now fine-tune our system so that the different fields $\phi_a$ all propagate at the same speed, that is all the matrices $G_a$ are equal, we may define $ g^{ij} = \frac{G_a^{ij}}{(\det (-G_a))^\frac{1}{D-1}}$. If we additionally define $g_{ij}$ as the inverse of $g^{ij}$, $g= \det(-g_{ij})$, $m_a= \frac{M_a}{g^\quarter}$ and we redefine the potential $V(\phi_a)\rightarrow \frac{1}{\sqrt{g}}V(\phi_a)$ we may rewrite the Lagrangian of \rf{genlag} as
\begin{align}
 \Lag(x) = -\sqrt{g}\left\{ \left(\sum_a \half g^{ij} \deli \phi_a(x) \delj \phi_a(x) + \half m_a^2\phi_a^2(x)\right) -V(\phi_a(x)) \right\}. \label{newlag}
\end{align}
Note that since all the $G_a$ had Minkowskian signature, $\gij$ will also have the same Minkowskian signature. Written in this way, the Lagrangian looks a lot like the Lagrangian of fields in a curved spacetime with metric $g_{ij}$. The only difference being that here $\gij$ is a constant independent of $x$. However, $\gij$ being a collection of coupling constants, nothing stops of from tuning those coupling constants locally to make them depend on $x$, the spacetime location. If we do that we obtain a new Lagrangian
\begin{align}
 \Lag(x) = -\sqrt{g(x)}\left\{ \left(\sum_a \half g^{ij}(x) \deli \phi_a(x) \delj \phi_a(x) + \half m_a^2\phi_a^2(x)\right) -V(\phi_a(x)) \right\}. \label{newworld}
\end{align}
which this time is identical to the Lagrangian of fields in a curved spacetime defined by the Minkowskian metric $\gij(x)$. 

Now we see how the conformal factor of the metric is set by the size of quantum fluctuations. If we multiply the metric by a conformal factor $\Phi^2 $ (i.e. $\gij\rightarrow \Phi^2 \gij$) then the terms in the Lagrangian will be multiplied by $\Phi^{2D}$ or $\Phi^{2(D+1)}$. Since, in quantum physics the amplitudes are given by a sum of $e^{\frac{1}{\hbar}\int d^{D+1}\Lag}$, if the Lagrangian gains a factor of $\Phi^{2D}$ it is equivalent to having $\hbar\rightarrow \frac{\hbar}{\Phi^{2D}}$. 

One interesting point to note is that the one thing which really have no choice about is the signature of the metric. If we start out with a Hamiltonian which is bounded from below\footnote{We talk here of the Bosonic case, the Fermionic case is of course more complicated.}, then the signature of the effective metric must be Minkowskian. This is very intriguing and might be telling us about some deeper more intricate relation between quantum physics and relativity. This, especially considering that, as previously mentioned, it is the size of quantum fluctuations which determine the conformal factor of the metric. In the following section we give a concrete realization of U(1) Yang-Mills on a Schwarzschild black hole background. 

\section{QED in Schwarzschild Spacetime}\label{QEDBH}
In \cite{WenLab} Wen proposed a concrete realization of his string-net condensate model for U(1) gauge Yang-Mills.  In this section we will show how one can take this in-lab condensed matter system of emergent light and locally tune the coupling constants of the system in order to end up with emergent light on an effective Schwarzschild black hole background. Thus creating a quantum system of a black hole with electromagnetic radiation. In addition to the interest of being able to observe a purely quantum gravitational effect, because the underlying quantum system will consist of a spin lattice, we will be implementing a minimum length scale. That is interesting because there has been some questioning \cite{transplanckjac} as to whether Hawking radiation would exist if there was a minimal length scale due to the fundamental way in which the continuum plays a roll in deriving the Hawking radiation\footnote{One requires the existence of trans-Planckian modes in the original derivation.}.  
 
The conceptual idea behind string-net condensate models is as follows (see \cite{WenLab,LRB1} for technical details). The underlying quantum system is a spin lattice or quantum rotor lattice. That is, a lattice, with on each edge, a quantum rotor or a spin. A string operator corresponding to a path $\gamma$ on the lattice corresponds to the ordered product of alternating raising and lowering operators along the path. The typical string-net condensate Hamiltonian then consists of three terms: a string tension term ($\propto \sum_k (S_k^z)^2$ in our spin $J$ example), a string fluctuation term ($\propto \sum_{\gamma} S_{\gamma}+S_{\gamma}^{\dag}$, the sum of all string of length 2 operators), and a Gau\ss{} constraint term ($\propto \sum_v \left(\sum_{e \ni v} S_e^z\right)^2$. When the coupling constant in front of the Gau\ss{} constraint is much bigger than the other two coupling constants, we effectively project down on to the sub-Hilbert space where the constraint is imposed exactly and thus there are no open strings. In particular, the the string fluctuation term for the effective Hamiltonian becomes a sum over plaquettes $P$ of the closed-string operators around the plaquettes\cite{WenLab}. In other words 
 $H_{\text{eff}} = J  \sum_e (S_e^z)^2 - \half g \sum_P (W_P + W_P\dagg)$,
where the first sum is over the edges of the lattice and $W_P$ is the closed-string operator over plaquette $P$. If, on the edges of the lattice, we have quantum rotors $\theta_e$ then $S_e^z = -i\pdif{}{\theta_e}$, $a_e = S e^{-i\theta_e}$ ($S\in\half\N$)and the continuum limit of the effective theory, when $J\ll g$, is $U(1)$-Yang-Mills, in $2+1$ dimensions with continuum Lagrangian
\begin{align}
\Lag_{2+1} =  \frac{\vect{E}^2}{J} - l^2 g B^2 = -\sqrt{\det(g_{2D})} g_{2D}^{ik} g_{2D}^{jm}F_{ij}(A)F_{km} ,\label{Lconsimple} 
\end{align}
where $l$ is the size of the lattice spacing and the coupling constants are given modulo constants of order one which depend on the exact lattice configuration.The metric is $g_{2D}:= -J^2 dt^2+ \frac{J}{g l^2}(dx^2 + dy^2) $. This Lagrangian is arrived at by defining the gauge field with support on lattice edges $\vect{A_e} := \theta_e \vect{e}$, where $\vect{e}$ is the unit vector pointing in the direction of edge $e$ and then taking the continuum limit.

It is also possible to make a 3+1 dimensional system of emergent light \cite{WenLab} by layering $2D$ lattices together so make a $3D$ lattice. In which case we obtain the following continuum Lagrangian, written in orthonormal coordinates, and with the third coordinate perpendicular to the layering
\begin{align}
\Lag_{3+1} =  \frac{1}{\zeta l J}\left(E_1^2 + E_2^2 +\zeta^2 E_3^2\right) - \frac{g l}{\zeta} \left(\zeta^2 B_1^2 +\zeta^2 B_2^2 + B_3^2\right), \label{L3consim} 
\end{align}
with $\zeta$ being the ration between the layer spacing and the lattice length $l$. If we now stack shells of spherical layers together, we obtain a Lagrangian of the form \rf{Lconsimple} with the metric
\begin{align}
  g_{3D} := \exp(2\Theta) \left\{ -dt^2 + \frac{dr^2}{9Jgl^2 \zeta^2}+ \frac{r^2}{9 Jgl^2} d^2\Omega\right\} , \label{echtemetrik3d}
\end{align}
if the condition $ 3 g = 16 J$, which ensures consistency of the metrical interpretation, is satisfied. We obtain an arbitrary conformal factor $\exp(2\Theta)$ because in $3+1$ dimensions, the conformal factor of the metric in the action of Yang-Mills exactly simplifies and does not appear in the action. $r$ is the radial distance and $d^2\Omega$ is the standard metric on the 2-sphere. Satisfying the metricity constraint, we automatically have that at cold enough temperatures, the system will be in the stringnet condensate phase of emergent light\cite{WenLab}. So we will have $U(1)$-Yang-Mills on the curved background. If we now wish that background to be Schwarzschild, we must impose
\begin{eqnarray}
 \exp(2\Theta) & = 9 J g l^2 = \left(1-\frac{2M}{r}\right)c^2\nn
\zeta & = c \left(1-\frac{2M}{r}\right) \label{solschwarz} , 
\end{eqnarray}
where $M$ is the mass of the black hole and $c$ is the desired speed of light. Since $\zeta$,$g$,$J$, and $l$ must all be positive quantities, we see that we can build the black hole only down to the horizon, we cannot cross the horizon as this requires a sign change in the metric. We can thus build the Schwarzschild spacetime from infinity down to the horizon by appropriately tuning $g$ and $J$ (which here vary between $0$ and a value of order $1$ and is thus not problematic) and piling up the layers closer and closer together close to the horizon. For a small enough black hole, we can build the metric to within an arbitrarily small fraction of the effective Planck length of the horizon: to get to within $\epsilon$ of the horizon, the smallest ratio we need between the inter-layer distance and the intra-layer lattice distance is $\zeta= \frac{c \epsilon }{2M+\epsilon}$.   
\newline

\bibliographystyle{iopart-num}
\bibliography{bib}
\end{document}